\documentclass[twocolumn]{aastex631}

\usepackage{amsmath}
\let\tablenum\undefined  
\usepackage{siunitx}
\usepackage{graphicx}  
\usepackage[nolist,nohyperlinks]{acronym}

\newacro{CHIME}{the Canadian Hydrogen Intensity Mapping Experiment}
\newacro{ISM}{interstellar medium}
\newacro{VLBI}{very long baseline interferometry}
\newacro{VDIF}{VLBI Data Interchange Format}
\newacro{DM}{dispersion measure}
\newacro{ESE}{Extreme Scattering Event}
\newacro{RFI}{radio-frequency interference}

\newcommand\review[1]{#1}

\begin{document}

\title{Echoes in Different Tempo: Long-Term Monitoring of Crab Echoes with CHIME}

\shorttitle{Echoes in Different Tempo}
\correspondingauthor{Thierry Serafin Nadeau}
\email{serafinnadeau@astro.utoronto.ca}

\shortauthors{Serafin Nadeau et al.}

\author[0009-0005-6505-3773]{Thierry Serafin Nadeau}
\affil{David A. Dunlap Department of Astronomy and Astrophysics, University of Toronto, 50 St. George Street, Toronto, ON M5S 3H4, Canada}
\affil{Canadian Institute for Theoretical Astrophysics, 60 St. George Street, Toronto, ON M5S 3H8, Canada}

\author[0000-0002-5830-8505]{Marten H. van Kerkwijk}
\affil{David A. Dunlap Department of Astronomy and Astrophysics, University of Toronto, 50 St. George Street, Toronto, ON M5S 3H4, Canada}

\author[0000-0001-5373-5914]{Jing Luo}
\affil{Canadian Institute for Theoretical Astrophysics, 60 St. George Street, Toronto, ON M5S 3H8, Canada}

\author[0000-0002-7164-9507]{Robert Main}
\affil{Department of Physics, McGill University, 3600 rue University, Montréal, QC H3A 2T8, Canada}
\affil{Trottier Space Institute, McGill University, 3550 rue University, Montréal, QC H3A 2A7, Canada}

\author[0000-0002-4279-6946]{Kiyoshi W. Masui}
\affil{MIT Kavli Institute for Astrophysics and
Space Research, Massachusetts Institute of Technology, Cambridge, MA, USA}
\affil{Department of Physics, Massachusetts Institute of Technology, Cambridge, MA, USA}

\author[0000-0002-2885-8485]{James W. McKee}
\affil{Department of Physics and Astronomy, Union College, Schenectady, NY 12308, USA}

\author[0000-0003-2155-9578]{Ue-Li Pen}
\affil{Institute of Astronomy and Astrophysics, Academia Sinica, Astronomy-Mathematics Building, No. 1, Sec. 4, Roosevelt Road, Taipei 10617, Taiwan}
\affil{Canadian Institute for Theoretical Astrophysics, 60 St. George Street, Toronto, ON M5S 3H8, Canada}
\affil{Canadian Institute for Advanced Research, 180 Dundas St West, Toronto, ON M5G 1Z8, Canada}
\affil{Dunlap Institute for Astronomy and Astrophysics, 50 St. George Street, University of Toronto, ON M5S 3H4, Canada}
\affil{Perimeter Institute of Theoretical Physics, 31 Caroline Street North, Waterloo, ON N2L 2Y5, Canada}

\begin{abstract}
    The Crab Pulsar is known to feature plasma lensing events known as echoes.
    These events \review{are characterized by} additional components in the pulse profile which are produced \review{by additional images formed when the pulsar's radio emission is deflected by ionized nebular material}
    \review{These components} are therefore delayed relative to the primary emission.
    We observed the Crab pulsar with \ac{CHIME} during its daily transits, creating an archive of baseband recordings of \review{giant pulses} in the 400$-800\,\unit{MHz}$ band.
    From these, we produced daily stacks of aligned pulses between late October 2021 and March 2024.
    We find that in these averages, echoes are readily visible throughout the observation period, and we identify clear groups of echoes with distinct behaviour in terms of their evolution with time and frequency.
    Many echoes exhibit dispersive delays consistent with being observed through excess column densities relative to the unscattered rays, but we also find two events where the dispersive delays indicate column density deficits.
    For the first time, we also find echoes for which the line of sight never directly intersects the intervening structures, resulting in events with non-zero minimum delays, of around ${0.5 \rm\,ms}$.
    The frequency and diversity of the observed echoes make the Crab an excellent target for long-term studies of astrophysical plasma lensing.
\end{abstract}


\section{Introduction} \label{sec:intro}

    Studies of radio sources and non-terrestrial ionized material have long been intertwined.
    It was through early studies of radio galaxies that scintillation was identified, first in the ionosphere and then in the interplanetary medium \citep{hewish75}.
    Since scintillation only occurs if the source is not resolved by the scintillation screen, its presence allowed the angular diameters of quasars to be constrained \citep{cohen67, little68}.

    The pursuit of scintillation studies of quasars in turn led to the discovery of pulsars \citep{hewish68}.
    Shortly afterwards, slow variations of the observed radio pulse intensities were identified and attributed to interstellar scintillation \citep{scheuer68, rickett69}, a process where the pulsar's radio emission interacts with the ionized plasma close to the line of sight as it travels to the observer, leading it to follow multiple paths which interfere with each other.

    In scintillation, localized changes in electron number density in the \ac{ISM} and the resulting variations in refractive index typically appear to lead to many different paths.
    However, electron density variations can also lead to more coherent lensing.
    If this lensing dominates, it can lead to what are known as \acp{ESE}, where the brightness of a quasar \citep{fiedler87, clegg98} or pulsar \citep{cognard93} varies dramatically in a pattern highly suggestive of a diverging and thus overdense plasma lens, relative to the ISM.

    Less dominant lensing can lead to ``echoes'' in pulse profiles, additional \review{images of the pulsar seen as extra} components \review{in the pulse profile} which persist over the course of many days, during which their delay relative to the usual emission components evolves parabolically, either receding from or approaching the main component.
    Such echoes were first identified in the Crab pulsar, after a bright event in 1997 which lasted for close to 100 days and was caused by the line of sight approaching and crossing a structure within the Crab Nebula \citep{smith00, backer00, lyne01}.
    Similar events from 1974 \citep{lyne75}, 1992, 1994 \citep{lyne01} and 1996 \citep{crossley07} were identified retrospectively, and more recent ones have been sporadically reported since \citep{driessen19,rebecca23a,rebecca23b}, though none of these were as prominent in duration or brightness as the 1997 event.
    Echoes have also been seen in a few other sources, such as PSR B1508+55 \citep{wucknitz18, bansal20} and PSR B2217+47 \citep{michilli18b}, though the responsible structures have been inferred to be interstellar in origin rather than local to these sources.

    While \acp{ESE} and echoes are due to discrete structures, it is less clear what causes scintillation.
    Originally, scintillation was thought to be the result of homogeneous and isotropic Kolmogorov turbulence in the \ac{ISM}, with the bending of light due to diffraction by very small structures \citep{rickett90}.
    More recent observations, however, suggest that it is instead dominated by localized and highly anisotropic ``thin screens'' \citep{putney06, rickett09, brisken10, stinebring22}.
    One model posits that these are thin under- or overdense sheets seen at grazing incidence, with corrugations in these sheets leading to lensing and the formation of multiple images \citep{goldreich06,pen14}.
    Hence, scintillation, \acp{ESE} and echoes may all be related \citep{romani87, jow24}, and indeed \ac{ESE}-like structures have now been inferred in pulsar scintillation patterns \citep{zhu23}.

    Given the above, stronger observational constraints on lensing structures would be very valuable.
    Independent of the detailed model, the observed signal can be described as the convolution of the intrinsic emission with an impulse response function, typically assumed to be a one-sided exponential \citep[see][]{williamson72, williamson73, williamson74},  that represents the effects of the interaction with the intervening material.
    The interaction would thus be probed best by a source that repeatedly emits impulses, which can be approximated as delta functions.
    Fortunately, close approximations to such sources exist, although only a handful are known \citep{kuzmin07}: pulsars which emit giant pulses.
    These giant pulses are extremely narrow, yet very bright, and thus close to proper impulses.
    Indeed, for two millisecond pulsars, it was possible to use giant pulses to retrieve the complex impulse response function \citep{main17, mahajan23}.

    The Crab pulsar emits such strong giant pulses that they dominate the average pulse profile.
    While these do not allow one to retrieve the complex impulse response---they last a few microseconds and are comprised of nanosecond-long ``nano-shots'' \citep{hankins03}, which are resolved by the Crab Nebula \citep{main21,rebecca23a}---they still provide a nearly ideal way to study echoes.
    To produce accurate profiles, however, one should take into account that the giant pulses do not occur every phase rotation, and that they occur randomly, or jitter, within narrow phase windows \citep{lundgren95, sallmen99, bhat08}.
    Hence, rather than create an average pulse profile, one should detect individual giant pulses and align then before averaging.

    The first detailed study using stacked giant pulses was done by \cite{serafinnadeau24}, using an archive of single pulses from the 42 foot telescope at Jodrell Bank.
    In this data set, they identified a large number of echoes, showing that these are readily visible in carefully constructed pulse stacks, and far more common than the prior sporadic detections had implied.

    The large number of observed echoes, and the fact that all observed echoes approached zero delay are inconsistent with prior models, which were tuned to the 1997 event and developed under the assumption that echoes are rare \citep{backer00, grahamsmith11}.
    Instead, \cite{serafinnadeau24} suggest sheet-like structures within the nebula seen at grazing incidence.
    They associate these with the thin ionized skins of elongated filaments, which would be seen at grazing incidence when the line of sight enters or exits a filament.

    Here, we expand on the observations and methods from \citet{serafinnadeau24}, using \ac{CHIME} to produce an archive of daily Crab giant pulse profiles from late 2021 to early 2024.
    In Section \ref{sec:data}, we describe our procedures to find and record the giant pulses, and in Section \ref{sec:analysis} how we combined them in daily stacks.
    We then present and discuss notable trends in groups of echoes in Section~\ref{sec:group_echo}, and in individual echoes in Section~\ref{sec:solo_echo}.
    Finally, we discuss ramifications of our results in Section~\ref{sec:ramifications}.
    Note that in our discussion we only make qualitative attempts to infer the structure of the actual lenses.
    In a companion paper \citep{serafinnadeau25b}, we show that one of the more isolated echoes identified here can be fairly well reproduced quantitatively within the picture of lensing in the ionized skin of elongated filaments.

\section{Data}\label{sec:data}

    The data for this paper were collected by \acf{CHIME}, a static interferometer composed of four identical $20 \times 100{\rm\,m}$ half-cylinders oriented along the North-South direction \citep{chime22}.
    Along each cylinder are 256 feeds sensitive to a frequency range of 400--800${\rm\,MHz}$, which collect polarization information along the North-South and East-West directions.
    \review{As the sky transits overhead, the signals from these feeds are digitized}, covering a field of view of more than 200 square degrees \citep{ng17}.

    The digitized voltages are beamformed using an ``FX'' correlator in various ways, including in 11 distinct tracking beams, which can simultaneously be pointed towards different regions of the sky.
    Ten of these flow into the \verb|CHIME/PSR| backend \citep{chime21}, and are typically recorded in fold or search mode, while the eleventh, dedicated to \ac{VLBI}, goes to a baseband recording backend \citep{cassanelli22}.

    We used the \ac{VLBI} beam to monitor the Crab during each of its daily transits.
    Below, we first describe the recording setup in some detail, as it has not been presented before, and then discuss the processing that was done on-site to detect and archive giant pulses.

    \subsection{Recording}\label{sec:data_record}

    The beamformed data from the \ac{VLBI} beam are sent to a dedicated machine, where they are written to storage.
    This allows for \review{the recording of many TB} of full-resolution data, which can either be processed on-site or transferred off-site to longer-term storage for future processing or analysis.
    The recording is done with a version of \verb|KOTEKAN| \citep{recnik15, renard21}, modified to write to eight separate $11{\rm\,TB}$ hard drives, each containing a $50{\rm\,MHz}$ subband of the total $400{\rm\,MHz}$ bandwidth.

    The split in subbands allows data to be read in a way that already partially accounts for dispersion, thus reducing the amount of data that needs to be kept in memory for dedispersion.
    For comparison, given the \ac{DM} of $\sim\!57{\rm\,pc/cm}^3$ of the Crab pulsar \citep{lyne93}, the dispersion delay between 800 and 400${\rm\,MHz}$ is 1.1 seconds, while it is only 0.31 seconds between 450 and 400${\rm\,MHz}$, overall resulting in an 8-fold reduction in memory usage during the dedispersion process.

    The \review{raw baseband }data \review{from each of these subbands }are recorded in \aclu{VDIF} \citep[\acs{VDIF}, ][]{whitney10}, where it is structured into files of 80 frames, each containing 625 samples of $2.56{\rm\,\unit{\micro\second}}$.
    Each sample contains the 128 frequency channels of a given subband -- one eighth of the 1024 channels the whole band is converted to by the correlator -- with for each channel the X and Y polarizations encoded each as 4+4 bit complex numbers (with X and Y corresponding to the East-West and North-South directions, respectively).

    With this system, we have been observing the Crab pulsar near daily since 2021 October 21, covering its transit through the \ac{CHIME} field of view with recordings lasting approximately 15 minutes (yielding about 680\,GB of data).
    In this paper, we focus on data gathered in the 896 days since the start, up to  2024 March 31.
    In total, this includes 830 recordings.

    \begin{figure*}
        \centering
        \includegraphics[width=\textwidth]{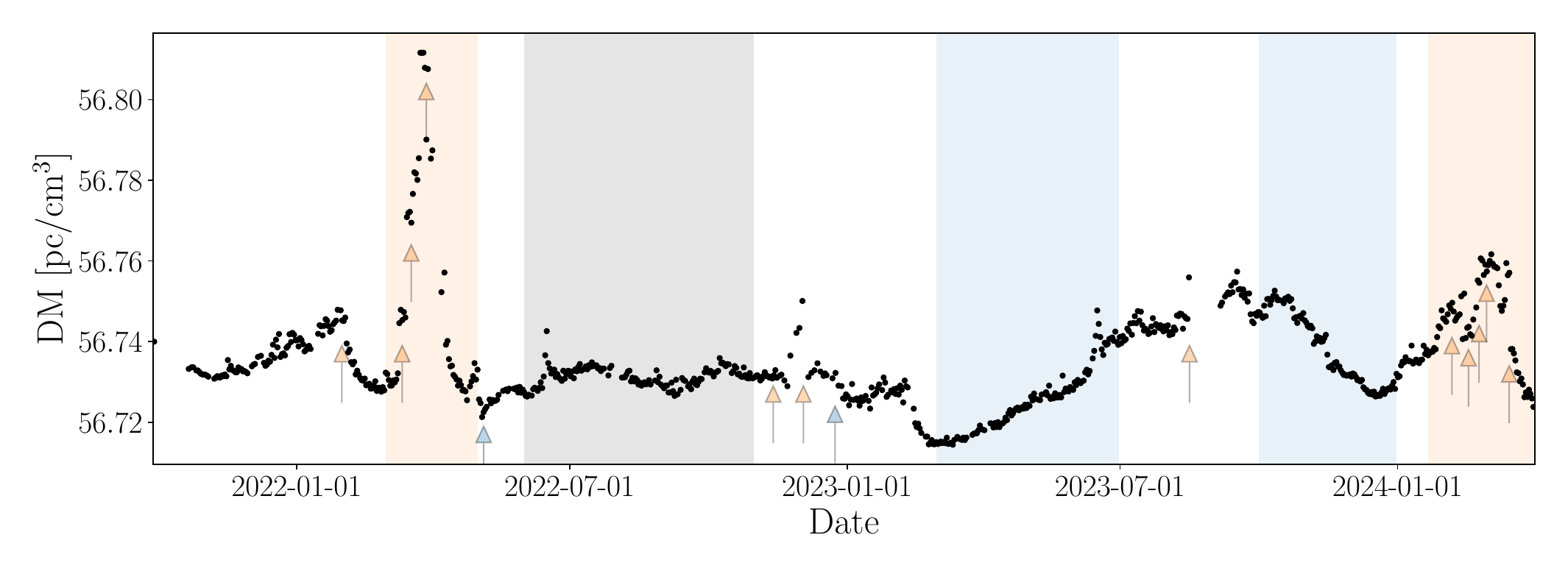}
        \caption{
            \Acf{DM} which maximizes the frequency-averaged peak signal-to-noise ratio for the brightest pulse on each day of observations, from 2021 October 21 to 2024 March 31.
            The shaded regions correspond to periods of high activity (orange; see Sect.~\ref{sec:active}), low activity with few echoes visible (blue; Sect.~\ref{sec:inactive}), and a period during which the echo curvatures are unusually low (grey; Sect.~\ref{sec:low_curv}).
            Arrows mark times that specific echoes appear: twelve echoes that become the dominant component of the profile, all of which have a frequency dependence in their delays consistent with excess \ac{DM} (orange tips, Sect.~\ref{sec:overdensity}), and two echoes that do not dominate the profile but stand out for showing delays consistent with \ac{DM} deficits (blue tips, Sect.~\ref{sec:underdensity}).
        }\label{fig:dm}
    \end{figure*}

    \subsection{Archiving}\label{sec:data_archive}

    Since the daily recordings correspond to 250\,TB of raw data each year, it is not sustainable to archive them all.
    Instead, we process the data on site to reduce the storage requirements.
    Inspired by the \verb|COBRA2| backend set up at Jodrell Bank \citep{shuxu14, mckee18, serafinnadeau24}, we create an archive of single pulses for each recording.

    The first step of the process starts immediately after the daily recording is complete, when we create a binned intensity stream, with a resolution of $256{\rm\,\unit{\micro\second}}$\review{, keeping full frequency resolution}.
    Here, we read the data one \ac{VDIF} file at a time, in order to reduce the memory impact, and avoid interfering with other recordings.
    This binned version is then incoherently dedispersed to a trial \ac{DM} of 56.74${\rm\,pc/cm}^3$, typical of the Crab pulsar \citep{lyne93}.
    Then, the dedispersed stream is summed over frequencies, and the peak with the highest signal-to-noise ratio identified.

    Next, the raw baseband data near the timestamp of this brightest peak is coherently dedispersed to the same trial \ac{DM}, relative to $800{\rm\,MHz}$ (the top of the band), with the \verb|baseband_tasks| Python package \citep{mhvk25}, again one $50{\rm\,MHz}$ subband at a time.
    From the resulting 8 dedispersed time streams, we extract 15000 samples ($38.4{\rm\,ms}$), starting from 2500 samples ($6.4{\rm\,ms}$) before the timestamp of the trigger and thus 12500 after ($32.0{\rm\,ms}$).
    These dedispersed data are then used to refine the dispersion measure, by identifying the \ac{DM} for which this trigger's peak signal-to-noise ratio is maximized after summing over the full frequency range.
    We show the inferred \ac{DM} values for the whole dataset in Figure~\ref{fig:dm}.

    Note that the above procedure can result in slight overestimates of the true \ac{DM}, since scattering causes the peak of the emission to shift to higher delays as the frequency decreases \citep{mckee18}.
    For our purposes of identifying giant pulses, however, it is the best estimate, since it ensures peaks stand out most relative to the noise.
    In principle, one could revise the \ac{DM} again after the fact, using all pulses, but we do not do that here.

    With the refined \ac{DM} in hand, we repeat the search in binned intensity, and now record timestamps for all peaks with a signal-to-noise ratio greater than~2.5.
    For the 512 brightest of these, we coherently dedisperse the raw baseband data to the same \ac{DM}, and again extract $38.4{\rm\,ms}$ (just over one spin cycle of the Crab Pulsar) of full-resolution data, starting $6.4{\rm\,ms}$ before each trigger's timestamp.
    These raw data are then saved, using 16-bit floating point numbers, and archived along with the original binned intensity stream.
    Overall, this process of only storing the brightest pulses reduces the total storage requirements by about an order of magnitude, from 680 to $67{\rm\,GB}$ per day.

    \begin{figure}
        \centering
        \includegraphics[width=\columnwidth]{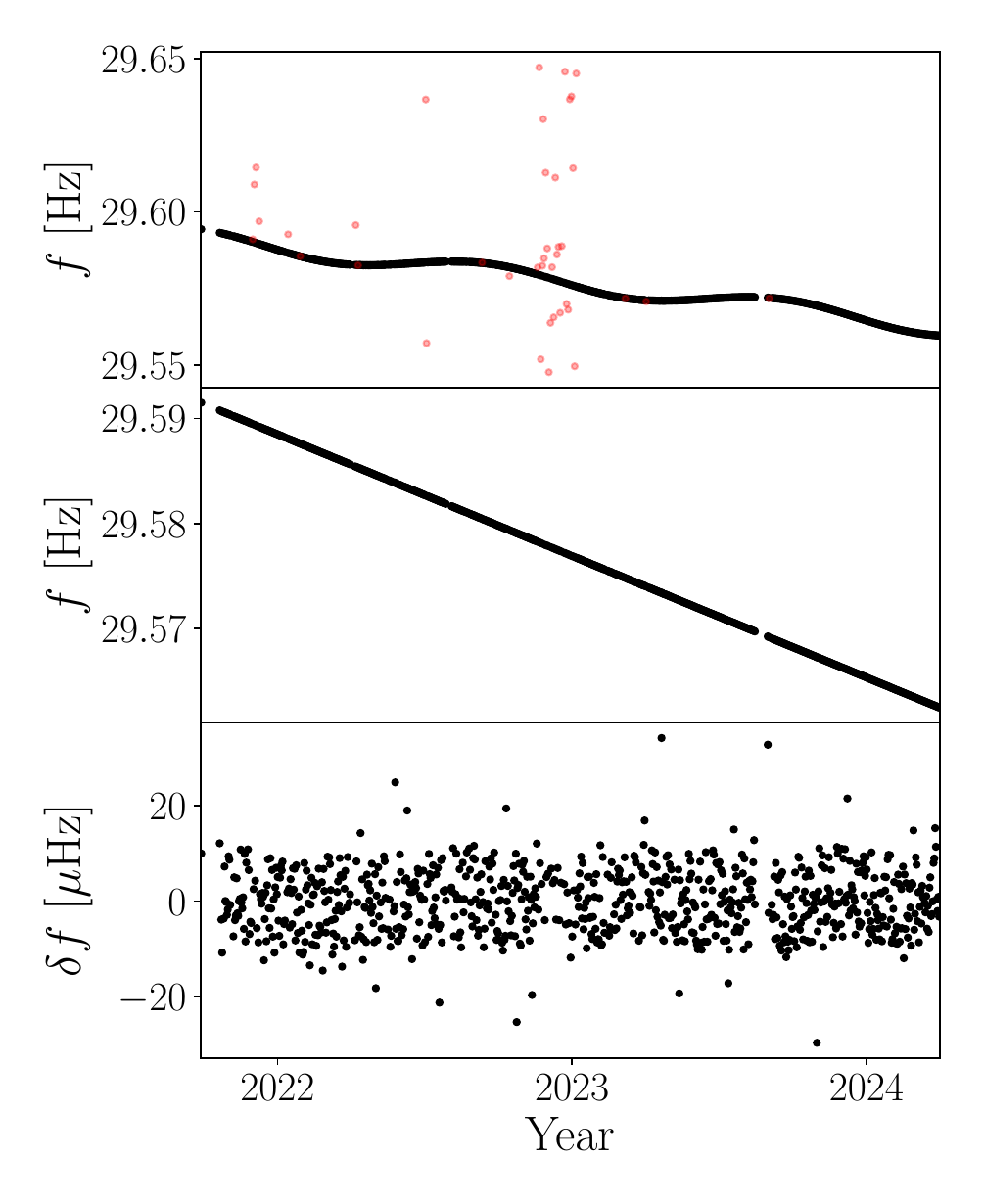}
        \caption{
            Spin frequencies inferred from our data, covering the period of 2021 October 21 to 2024 March 31.
            {\em Top:\/} Spin frequency inferred for each recording, with outliers \review{(from days with faulty recordings)} colored red.
            {\em Middle:\/} Same, but after removal of Doppler shifts due to Earth's motion.
            {\em Bottom:\/} Residuals relative to a third-degree polynomial.
        }\label{fig:spin}
    \end{figure}

    With the large number of timestamped triggers, we also produce daily estimates of the Crab pulsar's spin frequency.
    The main pulse and interpulse phase windows, where Crab giant pulses are known to occur, each occupy phase windows that are about $400{\rm\,\unit{\micro\second}}$ wide \citep{lundgren95, sallmen99, bhat08}.
    Starting from a typical spin frequency of $29.6{\rm\,Hz}$, the frequency which maximizes the number of triggers within the main pulse phase window is identified, to within $10{\rm\,\unit{\micro\hertz}}$, as shown in Figure~\ref{fig:spin}.
    This is then used to categorize archived triggers as main pulse, interpulse, or out of phase triggers.

    Of the 830 recordings, totaling 192 hours, 322670 saved triggers occurred during the main pulse phase window, 5482 during the interpulse phase window, and 29252 occur out of phase.
    These numbers show that our dataset is dominated by main pulses, reflecting that those are on average brighter than interpulses and are therefore preferentially included among the brightest triggers found each day.

\section{Processing}\label{sec:analysis}

    Prior observations of echoes relied mostly on folded profiles \citep{backer00, lyne01, wucknitz18, michilli18b, driessen19}, i.e., phase averages of the observed emission for which the noise is averaged out sufficiently that echoes are visible, or on very bright single pulses \citep{crossley07, rebecca23b}, where the pulse is bright enough that echoes peek through the noise floor.
    Both methods have disadvantages: bright single pulses grant clear snapshots of echoes, but there is no guarantee that sufficiently bright pulses occur often enough to obtain a good sampling of an echo throughout its evolution, while folded profiles allow sensitive measurements (even if sub-optimally since giant pulses are not produced every phase rotation), but the intrinsic jitter of the narrow giant pulses within their phase windows results in smearing of the averaged profile, including of any echoes.

    Following \cite{serafinnadeau24}, we avoid these issues by identifying the arrival time of individual giant pulses, and shifting them to a common pulse phase to produce daily averaged profiles.
    Below, we describe how we aligned the pulses and how we combined them into daily stacks.

    \subsection{Alignment}\label{sec:align}

    The dataset of \cite{serafinnadeau24} had an extremely small bandwidth, of only $5{\rm\,MHz}$ centered on $610{\rm\,MHz}$, which meant alignment could be done straightforwardly using the frequency-averaged pulses.
    In the eighty times wider \ac{CHIME} bandwidth, however, individual pulses exhibit a wide variety of chromatic behavior, ranging from different spectral slopes of the flux density to being constrained to a smaller subband or showing frequency dependent banding, which may even drift during the scattering tail \citep[for examples, see][]{bij21}.

\begin{figure*}
        \centering
        \includegraphics[width=\textwidth]{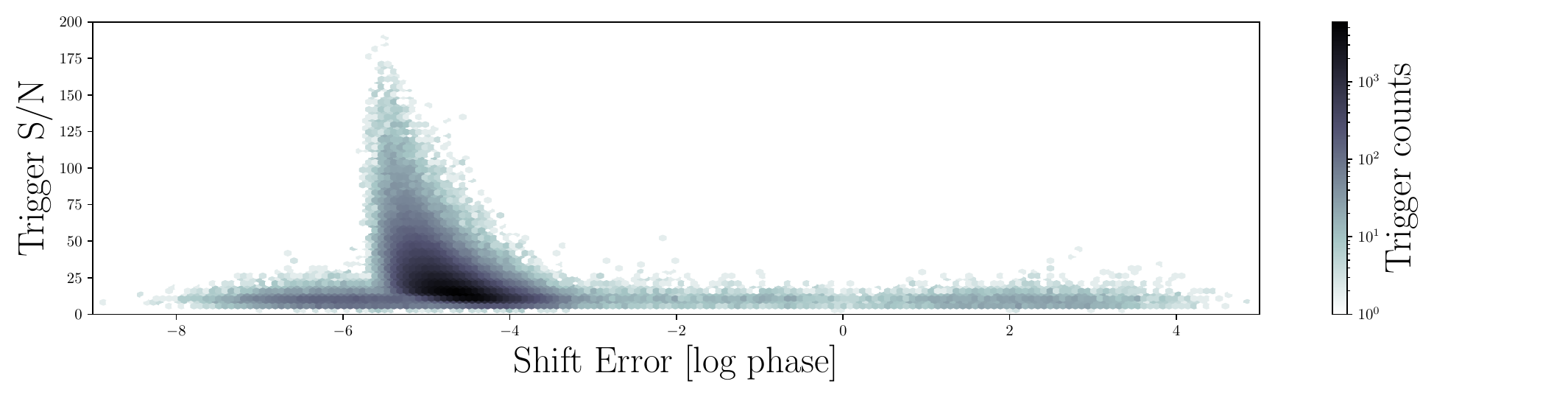}
        \caption{
          Distribution of uncertainties in the phase offset \review{(in units of cycle)} relative to the reference pulse for all triggers.
        }\label{fig:shifts}
    \end{figure*}

    Therefore, instead we correlate the individual triggers with a reference pulse, for which we take the brightest pulse of a given day.
    We correlate each channel separately, masking any channels with unusually low variance, typically due to over-saturation by \ac{RFI}.
    Then, the relative phase shift is determined by fitting the peak of the frequency-averaged correlation function with a Gaussian.
    As can be seen in Figure \ref{fig:shifts}, this allows us to measure the offset to very good precision, typically within single sample.

    \subsection{Stacking}

    \begin{figure*}
        \centering
        \includegraphics[width=0.9\textwidth]{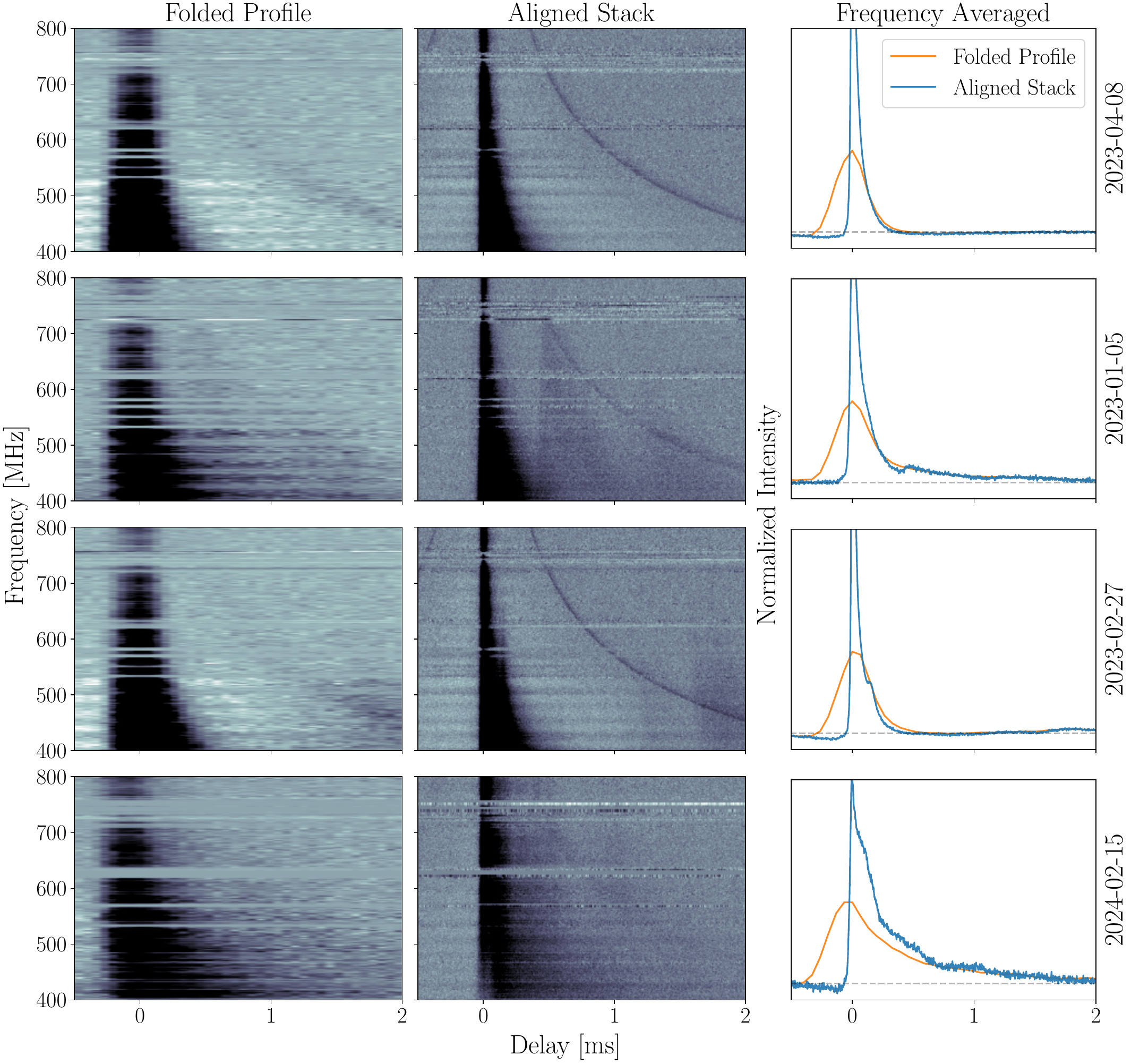}
        \caption{
            Comparison of folded profiles \textit{(left)} and aligned giant-pulse stacks \textit{(middle)} obtained with \ac{CHIME} on four days, as well as their frequency-averaged profiles \textit{(right)}.
            In the top row, profiles from 2023 April 8 are shown, when no echoes were visible.
            These illustrate the basic differences between the two methods of generating average profiles.
            The two middle rows show profiles from 2023 January 5 and 2023 February 27, when relatively clear echoes were present, at different delays from the main emission.
            The bottom row, from 2024 February 15, shows an example of less clear echoes.
            Note that the leading and lagging ``sweeps'' are artifacts from dedispersion in the presence of leakage between channels, while the apparent dimming in some narrow  frequency ranges is an artifact of storing data with a limited number of bits in the presence of strong signals.
        }\label{fig:stacks}
    \end{figure*}

    For creating the stacks, we first restricted the triggers \review{
    in two ways.
    First, we exclude any triggers which are not within 5\% of the main pulse or interpulse phase centers, as we do not expect to find giant pulses outside these phase regions.
    Secondly, we restrict triggers to those for which the fit to the correlation peak succeeded and had a shift uncertainty smaller than 0.01\% in phase (i.e., $\lesssim\!1$ time bin).
    This ensures that we do not contaminate the stack with pulses that were too low signal to be aligned properly, or with non-pulse triggers which are not related to the pulsar emission.}
    Altogether, this excluded 60655 of the 357404 total triggers, or about 17\%.

    The individual pulses were then shifted to the phase of the reference pulse (by \review{applying a phase rotation} in the Fourier domain\review{, corresponding to the desired shift in delay}), and added to the stack.
    We kept track of possibly masked channels (see above) and corrected for those in the average profile by dividing each channel by the number of times an unmasked channel was added to the stack.

    For radio observations of the Crab pulsar, the system noise is dominated by  emission of the surrounding nebula, so long as it is not resolved \citep{cordes04}.
    This is the case for our observations, as the \textrm{FWHM} of the \ac{CHIME} tracking beams is $\sim\!0.25^{\circ}$ at $800{\rm\,MHz}$ \citep{chime21}, while the nebula is only half this size along its major axis \citep{trimble68}.
    We use this to roughly remove the effect of possible gain variations in the aligned stacks, by dividing each frequency channel by its off-pulse median.

    We show four examples of aligned stacks in Figure \ref{fig:stacks}, comparing them to folded profiles produced by the \verb|CHIME/PSR| backend during the same days.
    As discussed above, one sees that the folded profiles are smeared.
    When clear echoes are observed in the stacks, these are only visible as moderate excesses in the scattering tail of the folded profiles if they have longer delays; those with shorter delays are lost in the folded profile.
    Similarly, echoes close to one another become indistinguishable, and if they are fainter, are subsumed into an apparently smooth scattering tail.

    \begin{figure*}
        \centering
        \includegraphics[width=\textwidth]{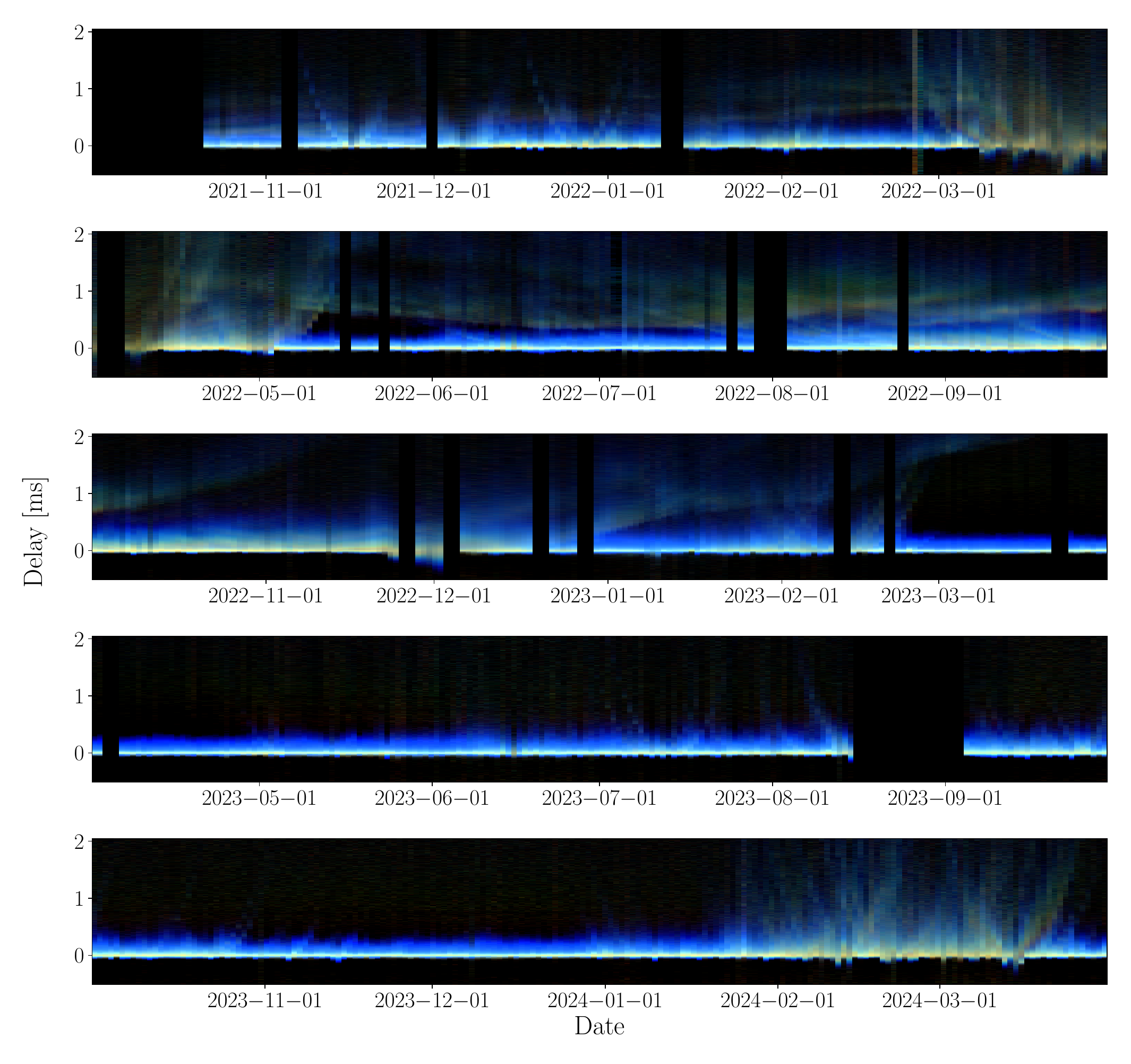}
        \caption{
            Daily stacks for the full period, from 2021 October 1 to 2024 March 31.
            \review{The colors shown are an RGB spectral composite, with power from different frequencies contributing different weights of red, green and blue. These weights increase and decrease linearly over 50MHz of the observation bandwidth, and plateau over 100MHz. For red, green and blue, these plateaus are centered at 700, 600 and 500MHz, respectively.}
            \review{As a result, red indicates there is only power at high frequencies, while blue indicates power restricted to low frequencies, and yellow-white indicates power is present across the whole bandwidth.}
            To avoid visual distraction, we mask bad days, and then interpolate over single-day gaps.
        }\label{fig:full_data}
    \end{figure*}

\section{Distinct Epochs in Echo Behaviour}\label{sec:group_echo}

    An overview of our daily profiles is shown in Figure~\ref{fig:full_data}.
    One sees that over the course of the observations, a wide variety of echo events are present.
    Individually, many of these offer little new substance to the overall discussion of echoes and their origins, other than by sheer numerical bulk, with similar cases having been previously outlined in prior work \citep{backer00, lyne01, crossley07, driessen19, rebecca23a, serafinnadeau24}.
    Instead, it is the behavior of collections of echoes over given time ranges that stands out the most.
    Below, we first separately discuss periods of high and low activity, and then discuss variations in how fast echoes evolve, focusing in particular on a period in which the evolution is very slow.

    \subsection{High Activity Periods}\label{sec:active}

\begin{figure*}
        \centering
        \includegraphics[width=\textwidth]{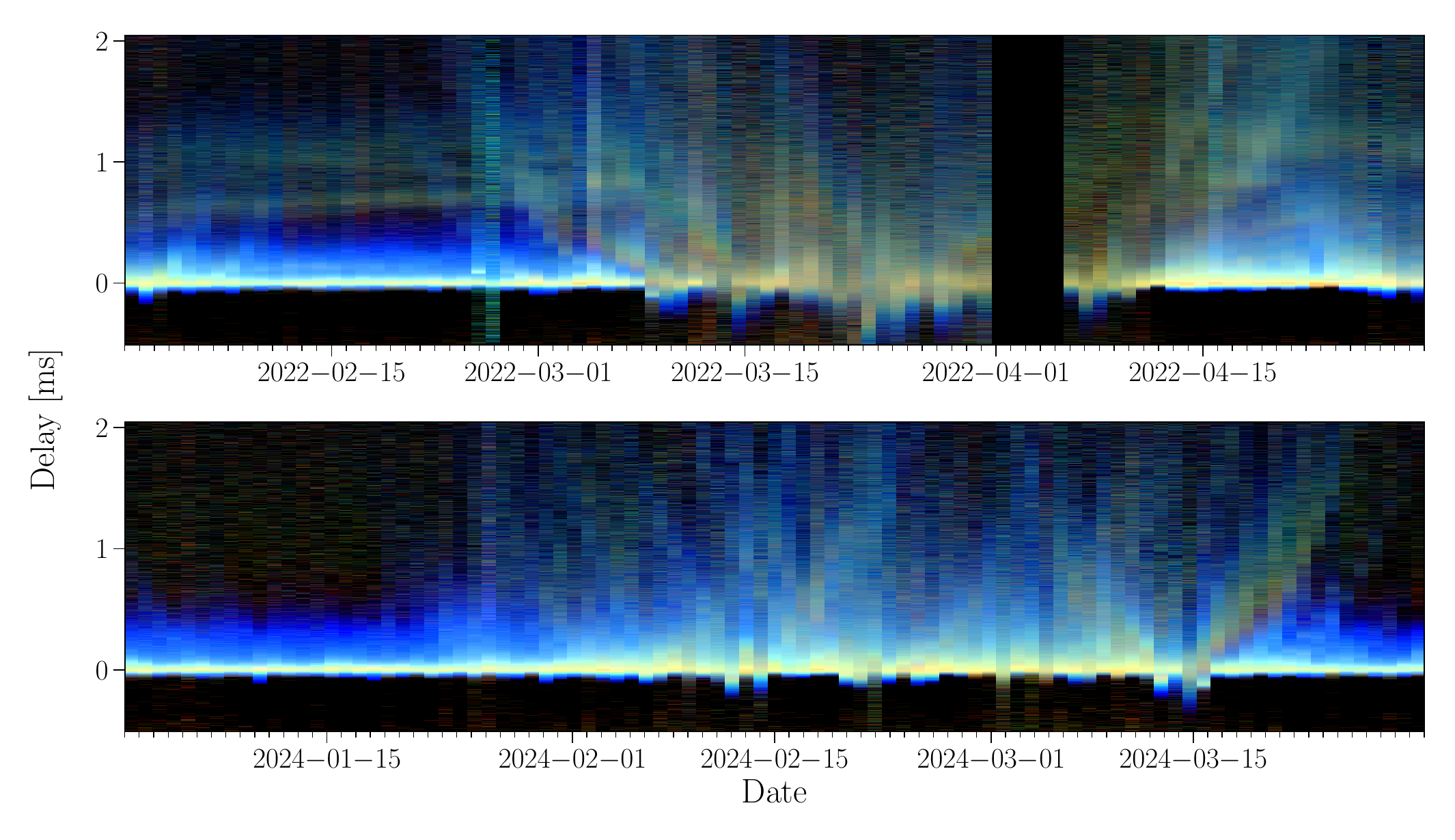}
        \caption{
            Daily stacks for two high activity periods, between 2022 February 1 and April 30 \textit{(top)} and between 2024 January 1 and March 31 \textit{(bottom)}.
            Representation and colors are as in Figure~\ref{fig:full_data}.
        }\label{fig:active}
    \end{figure*}

    By far the most prominent period of high echo activity occurred early in 2022 (see top panel of Figure~\ref{fig:active}), beginning in 2022 March with numerous echoes coming in, one after the other.
    On 2022 March 5, the first incoming echo of this series merges with the main component, and by March 10, an additional incoming echo component begins to dominate the profile as it approaches very low delay, while the original component remains visible.
    By March 14,  as yet another echo approached the line of sight, the original image fully disappears, and on March 15 the new arrival has replaced the prior echo as the dominant profile component.
    This pattern repeats once more on March 24, when another incoming echo becomes the dominant source of power in the Crab's profile.

    On 2022 March 31, a receding echo becomes visible, though its evolution is not well observed due to a system shutdown between April 2 to 5, after which the data were not good until April 13 as the gains had not been recalibrated properly.
    Nevertheless, another receding echo becomes visible on 2022 April 10, followed by a third on April 15.

    Throughout this event, \review{subsequent} incoming echoes \review{supplant} the previously dominant images, \review{and} the \review{observed signal strength progressively decreases, as the} view of the Crab \review{pulsar becomes increasingly} demagnified \review{by the intervening plasma.
    This demagnification then reverses} as echoes \review{begin} to recede from the line of sight through early April.
    In addition to this, the incoming images exhibited clear chromatic delays, consistent with different dispersion measures along the different images' light travel paths.
    Combined, they led to an increase in \ac{DM} of just over $0.08{\rm\,pc/cm^3}$ over the course of the event (see Figure~\ref{fig:dm}).
    This is the largest change in \ac{DM} reported for the Crab since the 1997 event \citep{lyne93, mckee18}, which showed a slightly larger jump of $0.12{\rm\,pc/cm^3}$ \citep{backer00}, from 56.8 to $56.92{\rm\,pc/cm^3}$.
    We discuss this jump as well as similar other ones in Section~\ref{sec:overdensity}.

    Another period of high activity occurred during the early months of 2024 (see the bottom panel of Figure~\ref{fig:active}).
    In this case, however, the total jump in dispersion and the demagnification of the pulsar emission are far less pronounced.
    Here, approaching and receding echoes are intermixed, with major incoming echoes merging with the main emission, or superseding it on 2024 February 6, February 12 and March 13, and major outgoing echoes appearing on February 14, February 17, February 24, and March 15.

    \subsection{Echo Desert}\label{sec:inactive}

\begin{figure*}
        \centering
        \includegraphics[width=\textwidth]{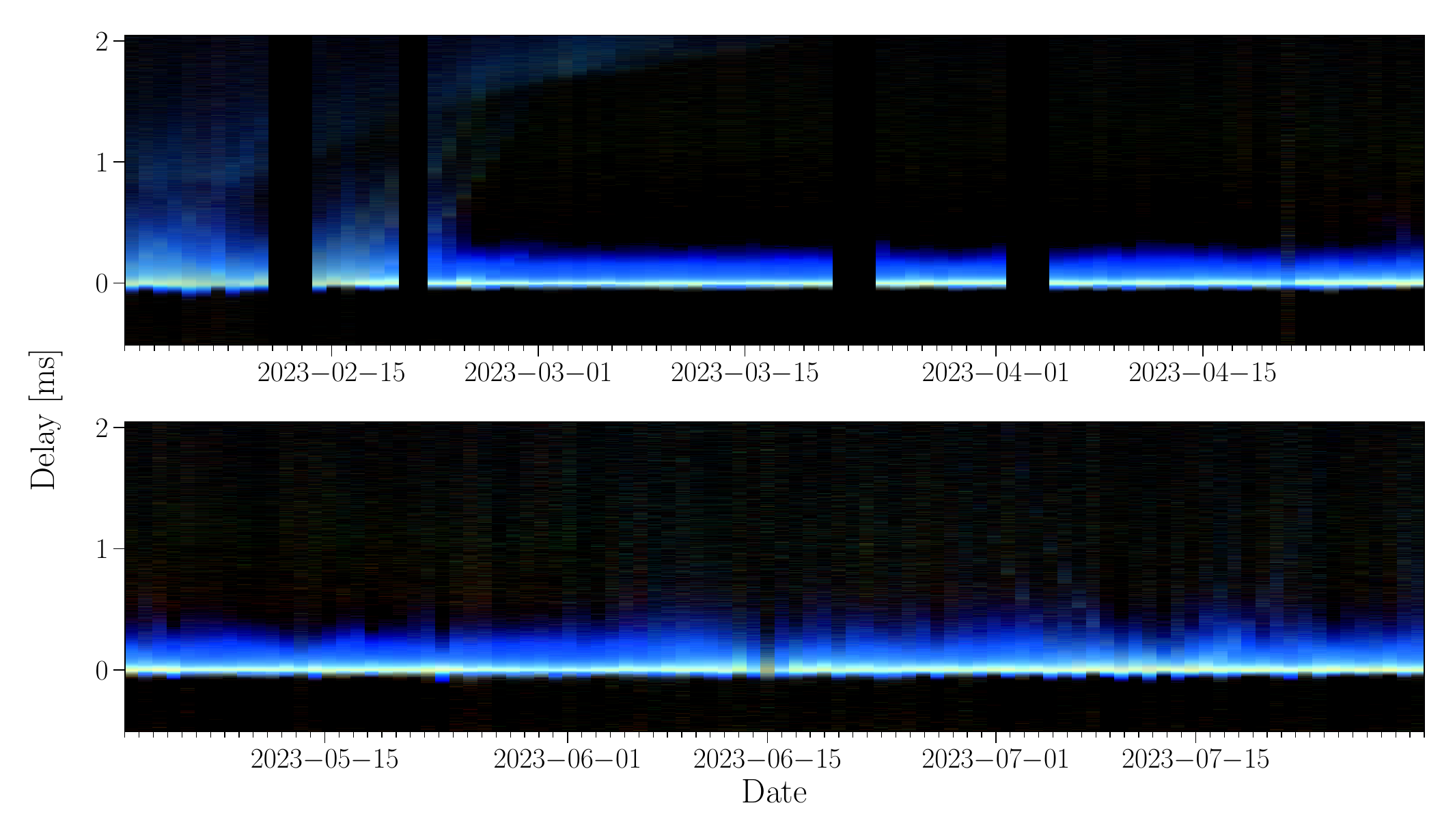}
        \caption{
            Daily stacks through the low activity period between 2023 February 1 and July 31, in two consecutive pieces.
            Representation and colors are as in Figure~\ref{fig:full_data}.
        }\label{fig:echoless}
    \end{figure*}

    In stark contrast to the above-described periods of high echo activity, during which echo images are constantly approaching and receding from the line of sight, there are also periods during which echoes are remarkably rare.
    In our dataset, this primarily occurs through the spring of 2023, as shown in the top panel of Figure \ref{fig:echoless}.
    During this period, the only \review{observed} echoes belong to a small cluster in early May.
    These are visible only at low frequencies, and fade out at very low delays\review{, and as a result do not show well on Figure \ref{fig:echoless}}.

    The transition into this low-activity period is very clear, \review{as seen to the left of the top panel of Figure \ref{fig:echoless}}, where excess power in the background recedes parabolically, leaving behind a low scattering environment.
    Conversely, the end of the period is subtle, as two echoes merge into the line of sight on 2023 July 7 and 11, respectively, before receding shortly thereafter.
    Multiple prominent echoes continue to be visible through the summer of 2023 (see Figure~\ref{fig:full_data}), until another, but shorter low activity period starts in 2023 November.

    During these two periods, the \ac{DM} variations are smooth and gradual, mostly increasing and decreasing, respectively (see blue shaded regions in Figure~\ref{fig:dm}).
    Given the lack of echoes, this is not surprising, as steep fluctuations in electron column density should lead to echoes.
    This logic is broken, however, by a small but distinct spike in dispersion over the days surrounding 2023 June 15, of $0.0013{\rm\,pc/cm^3}$, which is not associated with any echoes, and during which a demagnification of the pulsar is observed.
    This event is likely a result of the Crab's proximity to the Sun during this time, rather than due to nebular material, with a similar spike being visible in the same period during the previous year.

    In these periods with no echoes, the profile has what appears like an exponential tail.
    To get a quantitative sense, we fit the profile for a $5{\rm\,MHz}$ band around $610{\rm\,MHz}$ with a Gaussian convolved with an exponential.
    We find this gives a decent fit, with typical widths of $\sim\!6{\rm\,\unit{\micro\second}}$ and a scattering times of~$\sim\!26{\rm\,\unit{\micro\second}}$.
    The width is consistent with that found at much higher frequencies, of $1.7{\rm\,GHz}$, by \citet{rebecca23a}, giving confidence that this represents the average intrinsic width of the giant pulses.
    The scattering time is substantially lower than the minimum of $\sim\!0.1{\rm\,ms}$ found by \citet{mckee18} at 610\,MHz, but as discussed in footnote 15 in \cite{rebecca23a}, this likely reflects that those fits were done on the folded profile.

    The above scattering time is consistent with the lower range of scattering times seen at other frequencies.
    For instance, using the usual $\nu^{-4}$ scaling, our scattering time corresponds $\sim\!24{\rm\,ms}$ to $111{\rm\,MHz}$, at the low end of the range of 10--$115{\rm\,ms}$ seen by \cite{losovsky19}), while it would correspond to $0.12{\rm\,\unit{\micro\second}}$ at 2.3\,GHz, similar to the $0.07{\rm\,\unit{\micro\second}}$ inferred from the decorrelation bandwith of \cite{cordes04}, which according to \cite{rebecca23b} was very low compared to other high-frequency measurements.
    Likely, as suggested by \citet{rudnitskii16} and \citet{rebecca23b}, during these periods of low activity the scattering is dominated by the interstellar medium rather than the Crab Nebula.

    \begin{figure*}
        \centering
        \includegraphics[width=\textwidth]{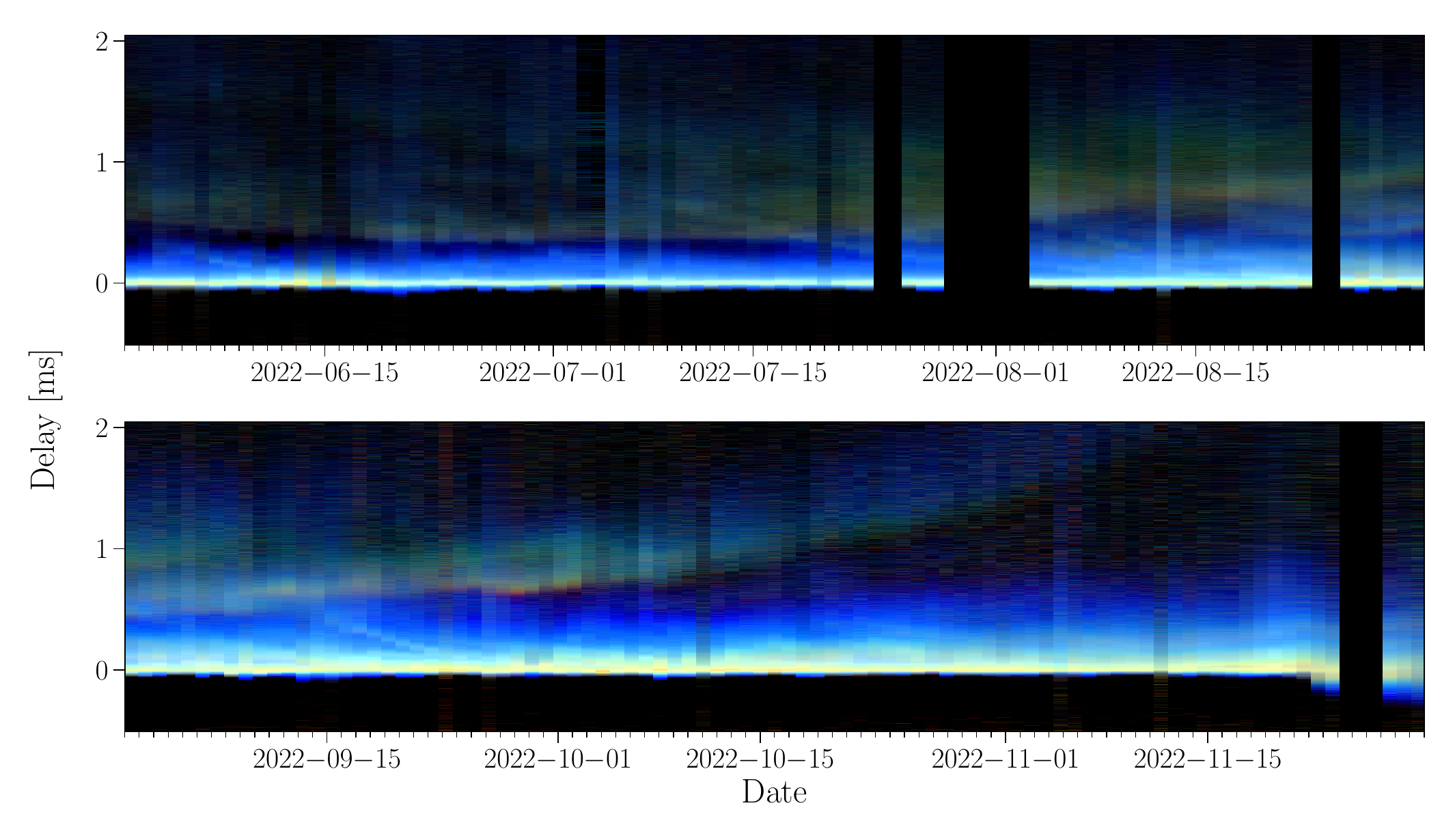}
        \caption{
            Daily stacks across the period between 2022 June 1 and November 30, when echoes evolved very slowly.
            Note that the panels here cover the same number of days as those in Figures~\ref{fig:active} and~\ref{fig:echoless}, in order to highlight the substantially lower curvature of the sharper echoes seen here.
            The whole period is shown again on a compressed scale in Figure~\ref{fig:nonzero}, to highlight the echoes that do not cross zero delay.
            Representation and colors are as in Figure~\ref{fig:full_data}.
        }\label{fig:low_curv}
    \end{figure*}

    \subsection{Echo Groups}\label{sec:low_curv}

    Between the different echoes, one notices that some evolve much faster than others (i.e, along steeper parabolae; see Figure~\ref{fig:full_data}).
    In the limit where the pulsar and screen distances from the observer are much larger than the distance between the pulsar and screen, $d_\textrm{ps}$, an echo's delay $\tau$ is given geometrically by,
    \begin{equation}\label{eqn:delay}
      \tau = \frac{1}{2c}\hat\alpha^2 d_{\rm ps} = \frac{x^2}{2cd_{\rm ps}},
    \end{equation}
    where $\hat\alpha$ is the angle by which radiation is bent, and where in the second equality we used $\hat\alpha=x/d_{\rm ps}$, with $x$ the distance between the pulsar and the lens projected on the sky.
    Writing $x=v_\textrm{ps} \cos\psi$, where $\Delta t$ is the time measured relative to when the pulsar and echo-producing structure line up relative to the observer, $v_{\rm ps}$ is the transverse velocity difference between the pulsar and the structure, and $\psi$ the angle between the velocity vector and the normal to the structure (see Figure~6 in \citealt{serafinnadeau24}), one finds that $\tau$ will evolve quadratically with time, as
    \begin{equation}\label{eqn:delay_evol}
      \tau = \frac{\left( v_\textrm{ps} \cos\psi \right)^2}{2cd_\textrm{ps}} \Delta t^2
      = \eta \Delta t^2,
    \end{equation}
    where we implicitly defined the curvature parameter~$\eta$.

    Scaling the velocity to the pulsar velocity relative to the center of the nebula, of $120\pm23{\rm\,km/s}$ \citep{kaplan08}, and the \review{pulsar-screen} distance to $1{\rm\,pc}$, within the range of 0.5--2\,pc inferred from optical emission lines \citep{martin21}, one expects curvatures around,
    \begin{equation}
      \eta = 5.8{\rm\,\unit{\micro\second}/day^2}\;\cos^2\psi
      \left(\frac{v_{\rm ps}}{120{\rm\,km/s}}\right)^2
      \left(\frac{d}{1{\rm\,pc}}\right)^{-1}.
    \end{equation}

    In our data set, the curvatures vary greatly, with echoes leading up to the high-activity period of 2022 March having the highest curvatures (and thus fastest evolution), of $\sim\!20{\rm\,\unit{\micro\second}/day^2}$, while those during the high-activity period have curvatures of $\sim\!7$ to $10{\rm\,\unit{\micro\second}/day^2}$.
    In the aftermath of the 2022 March event, echoes that evolve much more slowly are seen, with curvatures between 0.4 and $1.2{\rm\,\unit{\micro\second}/day^2}$, and echoes remain slow until the end of 2022 (see Figure \ref{fig:low_curv}).
    In early 2023, echoes evolve faster again, with curvatures near $10{\rm\,\unit{\micro\second}/day^2}$, until the low-activity periods described above begin.
    After the low-activity periods, echo curvatures range between 15 and $16{\rm\,\unit{\micro\second}/day^2}$ until the second high-activity period, where they hover around $10{\rm\,\unit{\micro\second}/day^2}$.

    The range in curvatures mostly overlaps with that seen in the prior study by \cite{serafinnadeau24}, where most identified echoes over a period spanning over 1700 days had curvatures ranging between 6 and $10{\rm\,\unit{\micro\second}/day^2}$, and a few exhibited larger curvatures, up to a maximum $17{\rm\,\unit{\micro\second}/day^2}$.
    These values reinforce the conclusion of \cite{serafinnadeau24} that most echoes likely originate near the inner part of the nebula seen in optical emission lines, at $\sim\!0.5{\rm\,pc}$ \citep{martin21}, and that the pulsar velocity is towards the upper range of what was inferred by \cite{kaplan08} and/or that the relative velocity is at times somewhat larger due to contributions from the \review{motion of the} screen.
    As echoes with the highest curvatures are those which are aligned perpendicular to the bulk motion ($\psi = 0$), we  update these constraints on the relative velocity to $v_{\rm ps} = 157 {\rm\,km/s}$ (Eq. \ref{eqn:delay_evol}).

    The period that is very different from any seen by \cite{serafinnadeau24} is that of extremely slowly evolving echoes that stretches through much of 2022 (see Figure~\ref{fig:low_curv}), which have curvatures even lower than the lowest curvature echo seen before, of $2{\rm\,\unit{\micro\second}/day^2}$ for the 1997 event.
    This is unlikely to reflect a low value of the relative velocity $v_{\rm ps}$: material along the line of sight to the center of the nebula moves mostly radially, so its transverse motion is unlikely to be large enough to nearly cancel out that of the pulsar.
    Similarly, the low curvature cannot be due to a large distance $d_{\rm ps}$: the  observations of line emission in the nebula constrain that to be smaller than $2{\rm\,pc}$ \citep{martin21}.
    Therefore, the low curvature must be the result of a relatively close alignment of the filamentary axis with the pulsar's path on the sky, so that $\cos\psi$ is small.

    The fact that echoes occur in groups, and that those within a group share similar curvature, suggests strongly that the structures responsible for a given cluster of echoes share the range of properties which define curvature.
    This would arise naturally if echoes are caused by substructures of larger nebular filaments, and that it is the properties of these filaments that determine the screen distances, orientations and velocities for a given cluster of echoes.

\section{Behaviour of Individual Echoes}\label{sec:solo_echo}

    The delays of most echoes do not appear to have large frequency dependence.
    In some cases however, the delays at low frequencies are significantly larger than at higher frequencies, suggesting extra dispersion, while in a few others, the opposite is seen.
    Below, we discuss these two cases first.
    Next, we discuss other unusual echoes, in particular some that approach but never reach zero delay, and others for which the dependence of their brightness with frequency seems unusual.

    \begin{figure}
        \centering
        \includegraphics[width=\columnwidth]{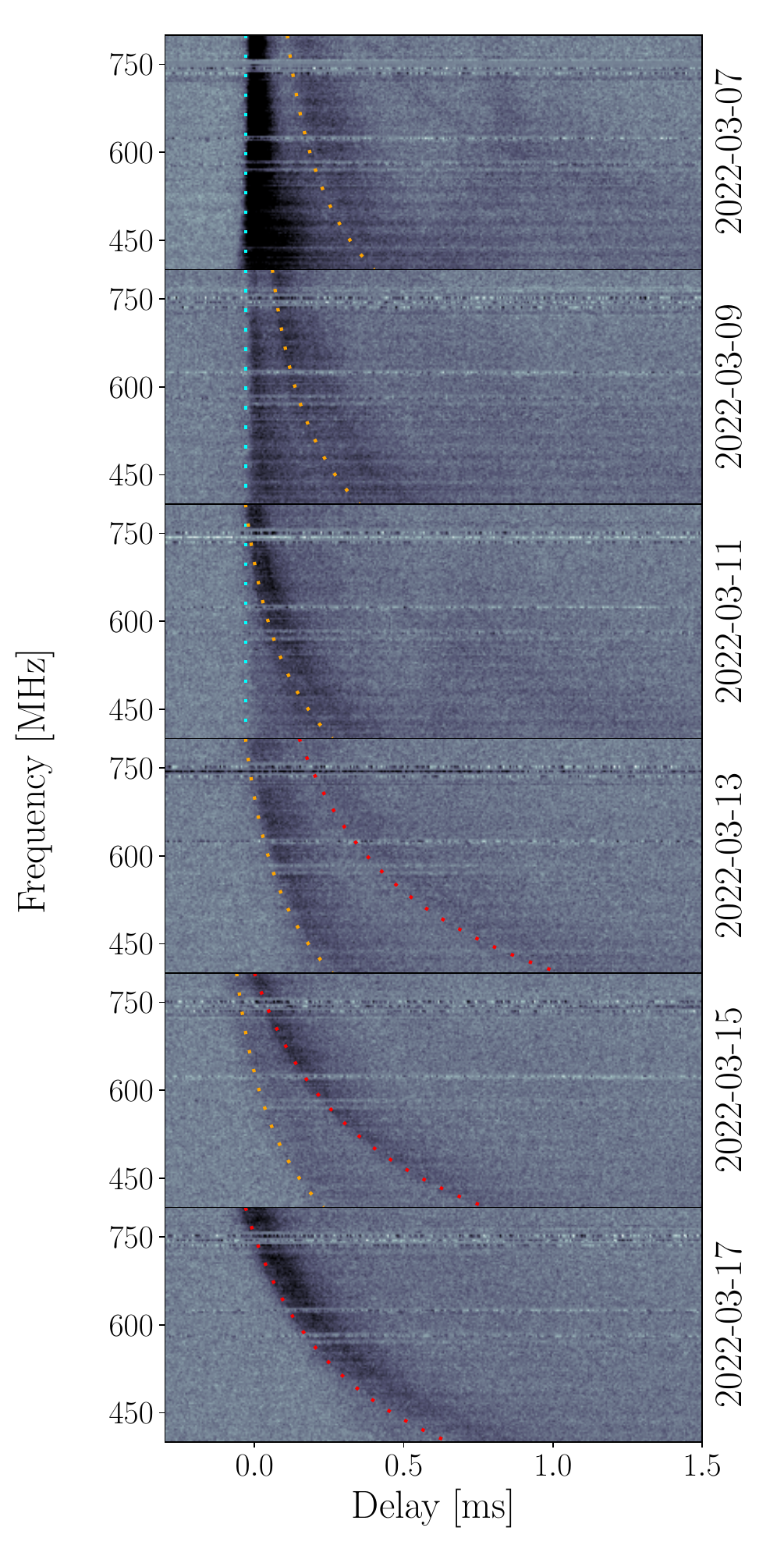}
        \caption{
            Daily stacks for every other day between 2022 March 7 and 17, all dedispersed to the same \acf{DM}, of $56.730{\rm\,pc/cm^3}$.
            The leading edge of the component which dominates the profile on March 7 {\em(top)} highlighted with a cyan dotted line, and the two echoes that come in over this period are similarly highlighted with orange and red dotted lines, respectively.
            For the first echo, the excess \ac{DM} does not change noticeably -- we used $0.015{\rm\,pc/cm^3}$ in all five panels for the orange dotted lines -- but for the second echo, which is seen over a larger range of delays, the excess clearly becomes smaller as it approaches zero delay -- the red dotted lines in the bottom three panels are drawn with excess \acp{DM} of $0.045$, $0.04$, and $0.035{\rm\,pc/cm^3}$, respectively.
        }\label{fig:overdense}
    \end{figure}

    \subsection{Excess Dispersion}\label{sec:overdensity}

    The echoes which provide the clearest evidence of extra dispersion, with the echo delay notably increasing with decreasing frequency, occur during the high-activity periods of 2022 March and 2024 February (though other instances are visible throughout the dataset, with the clearest one on 2022 November 28).
    We show individual time-frequency profiles from the 2022 March period in Figure~\ref{fig:overdense}.
    For this figure, to see changes in \ac{DM} more clearly, we dedispersed each day to the same \ac{DM} (the best-fit \ac{DM} found for the first day) rather than to the best-fit \ac{DM} of each individual day.
    One sees that the two echoes that come in during this period have increasingly larger \ac{DM}, i.e., the radiation travels through a larger electron column density.
    In detail, one sees that the excess \ac{DM} of the second echo, which dominates the profile at the end of the series, decreases as the echo approaches zero delay, by $\sim\!0.01{\rm\,pc/cm^3}$ over the four days  where it can be clearly demarcated (red dotted lines in the bottom three panels).

    That echoes in the Crab pulses would show different \ac{DM} is not surprising, given that they have long been thought to be the result of plasma lensing in the Crab Nebula \citep{backer00, grahamsmith11, serafinnadeau24}, which is expected to occur given the presence of the partially ionized filaments seen at optical wavelengths \citep{duncan21, osterbrock57, trimble68, henry82, sankrit98}.
    Such lensing requires the trajectories for the images to pass through different parts of the structures, which naturally leads to differences in dispersive delay.
    Indeed, dispersive delays were also inferred for the 1997 event, for which the \ac{DM} evolution also showed a jump as the echo became the dominant component, and at the time of the jump, the pulse profile components showed both the old and new \ac{DM} \citep{backer00}.

    The general behaviour can be understood by considering that the trajectory of any echo is bent in the lens, by an angle which depends on the gradient of the electron column density $\nabla_x N_e$ as,
    \begin{equation}\label{eqn:bending}
      \hat{\alpha} = \frac{c^2 r_e}{2 \pi \nu^2} \nabla_x N_e,
    \end{equation}
    where $r_e$ is the classical electron radius and $\nu$ the frequency.
    As the pulsar approaches the overdense structure and an echo first becomes visible, the echo's path necessarily has to go through the lens at the part with the largest gradient.
    Typically, for a smooth structure with a single bump in column density, the maximum gradient will occur on the side, where the column density is about half the maximum.
    Then, as the pulsar approaches more closely, the echo will \review{generally} split into two images, one each for the points of slightly lower $\nabla_xN_e$ on each side of its maximum.
    Again, \review{generally}, the images will initially have the same magnification, but the magnification of the image closer to the pulsar will increase as the separation decreases, until it begins to dominate.
    Since this brighter image goes through a lower total column density, one thus expects the excess \ac{DM} of an echo to be lower for lower delay \citep[see][for a more in-depth discussion]{serafinnadeau25b}.

    An important corollary to the above is that if one measures best-fit \acp{DM} for profiles, their evolution with time does not map linearly to changes in column density as a function of spatial position.
    Hence, lens models directly based on the observed \ac{DM} variations of a source can provide at best approximate descriptions of the physical structures involved.
    Nevertheless, the direction of change can be instructive: looking at Figure~\ref{fig:dm}, we see indeed that around the times discussed here, there are substantial increases in \ac{DM}, and when the overall \ac{DM} decreases, receding echoes with excess \ac{DM} are seen.

    In Eq.~\ref{eqn:bending}, one sees that the bending angle depends on frequency as well, and hence the trajectories for a given echo for different frequencies cross the lens at different points, and thus probe different excess \ac{DM} and have slightly different geometric delay.
    This was also noted by \cite{backer00}, who showed that the frequency-dependent component of the geometric time delay, which scales with $\nu^{-4}$, is expected to be small relative to the frequency-averaged dispersive delay.
    They were unable to test this, however, because of insufficient frequency sampling of the 1997 event.
    Our results here confirm that the frequency dependent delay in these higher \ac{DM} events is dominated by dispersive effects, well-matched to a $\nu^{-2}$ scaling.
    Unfortunately, it is difficult to quantify or constrain any $\nu^{-4}$ contributions from refractive effects, as the echoes are not that well defined; a better constraint would likely require detailed modeling of the temporal broadening of echoes.

    Finally, we note that Eq.~\ref{eqn:bending} also implies that the time that echoes are first visible should also depend on frequency.
    We return to this type of chromatic behaviour in Section~\ref{sec:chromatic}.

    \begin{figure}
        \centering
        \includegraphics[width=\columnwidth]{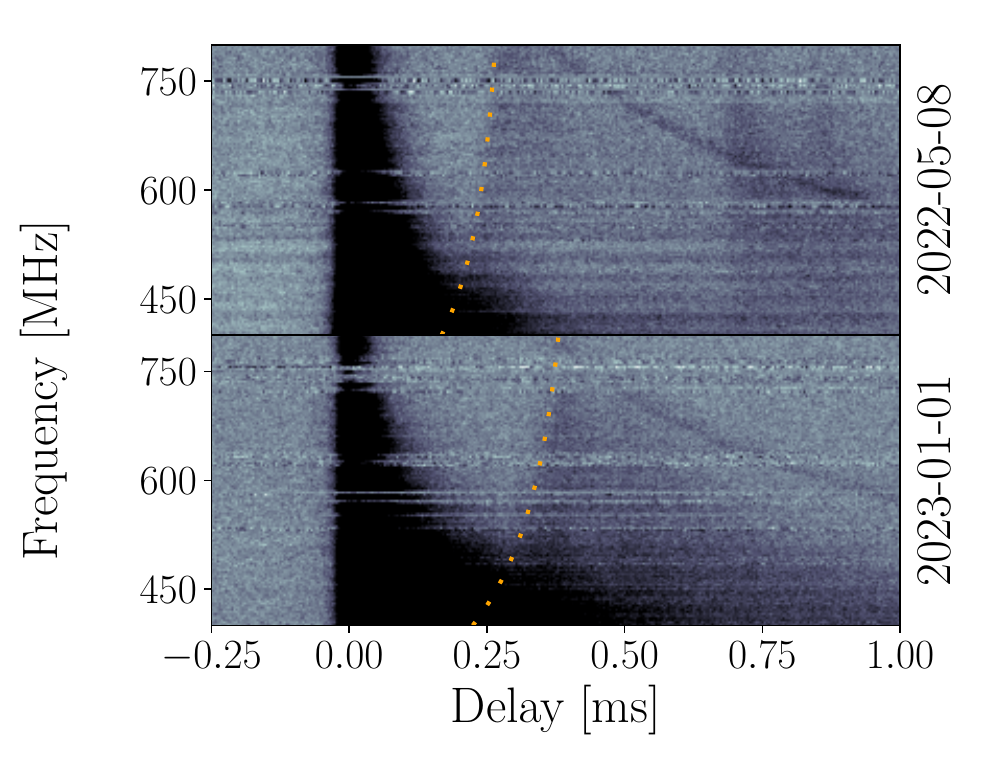}
        \caption{
            Daily stacks from 2022 May 8 \textit{(top)} and 2023 January 1 \textit{(bottom)}, showing echoes that appear to have a \ac{DM} deficit relative to the main component.
            The leading edges of the echoes are marked orange dotted lines, which assume \ac{DM} deficits of $-0.005$ and $-0.008{\rm\,pc/cm^3}$ in the top and bottom panel, respectively.
        }\label{fig:underdense}
    \end{figure}

    \subsection{Dispersion Deficits}\label{sec:underdensity}

    In contrast to the echoes which show the clear markings of overdensities, our dataset also includes two echoes that appear to have frequency dependencies in the opposite direction, with the observed delay being larger at higher frequencies.
    Snapshots of the two are shown in Figure \ref{fig:underdense}.
    The first occurs in early 2022 May, shortly after the end of the 2022 March high-activity period.
    This echo, which splits off from the main emission on May 6 and becomes too diffuse to reliably track after May 13, clearly shows a frequency dependence in its delay indicating a \ac{DM} deficit, of $-0.005{\rm\,pc/cm^3}$.
    The second echo, which first appears on 2022 December 30 and fades away on 2023 January 7, is more smeared than the first, but also appears to show a frequency-dependent delay indicative of a \ac{DM} deficit, of $-0.008{\rm\, pc/cm^3}$.

    Echoes with \ac{DM} deficits would naturally be produced if there is an underdense region, leading to electron column density deficit.
    For the first echo, there seems to be a corresponding dip in the observed \ac{DM} (see Figure~\ref{fig:dm}), but this is not the case for the second echo.
    As noted above, however, the \ac{DM} curve cannot be interpreted straightforwardly; indeed, we also see echoes that show excess \ac{DM} at times that nothing particularly interesting seems to happen in our best-fit \acp{DM}.

    Nevertheless, the echoes indicate that at times the line of sight crosses regions with reduced electron column density.
    As shown by \cite{serafinnadeau24}, given the densities inferred from the optical emission lines, the echoes require any overdense regions to be sheet-like, causing echoes only when seen at grazing incidence, when the total column density gradient becomes sufficiently large.
    Their logic would also apply to underdense regions: to produce echoes, those must also be sheetlike and seen nearly edge-on.
    While overdense sheets are fairly naturally associated with the ionized skins of the filaments, we cannot think of a natural physical origin for underdense sheets.
    Instead, it seems more likely that the dips in electron column density responsible for the underdense echoes occur because multiple overdense sheets appear close together on the sky, leaving an apparently underdense region in between.

    \begin{figure*}
        \centering
        \includegraphics[width=\textwidth]{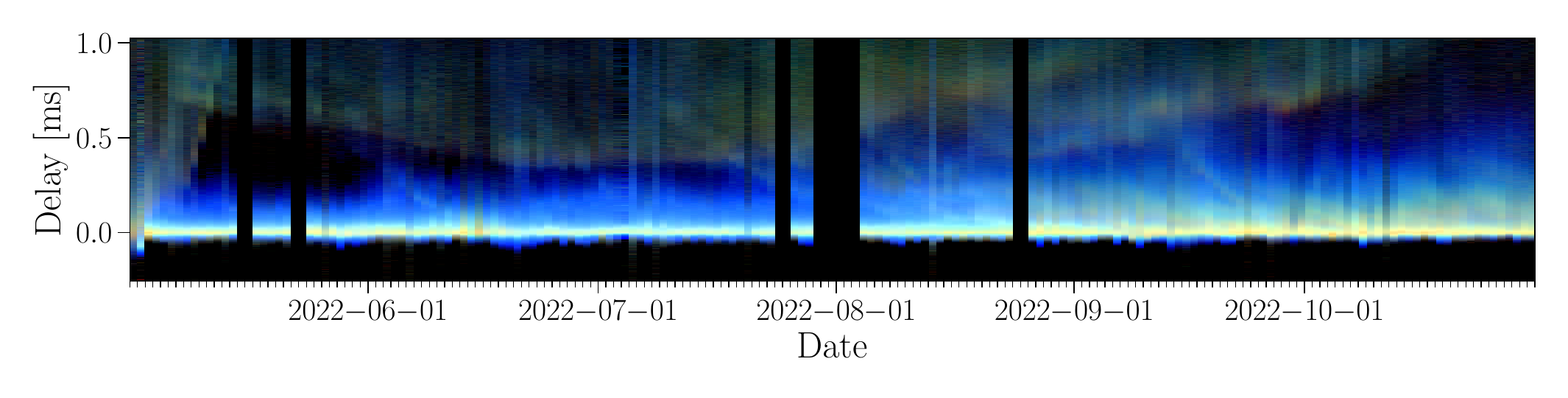}
        \caption{
            Period during which non-zero delay echoes are visible, between 2022 May 1 and October 31.
            This covers a similar period to that shown in Figure \ref{fig:low_curv}, but with an aspect ratio more suited to viewing the slowly evolving echoes that do not reach zero delay.
            Representation and colors are as in Figure~\ref{fig:full_data}.
        }\label{fig:nonzero}
    \end{figure*}

    \subsection{Non-Crossing Echoes}\label{sec:event_no-cross}

    Echoes of the Crab pulsar have now been studied for over two decades, if with irregular coverage.
    So far, all echoes were seen to either approach or recede from the leading component, i.e., they always appeared at some point to have zero delay.
    This demonstrated that the pulsar consistently passed behind the structures responsible for the echoes, rather than simply passing near them on the sky.
    Given this, so far it has only been possible to set lower limits to the sizes of the structures responsible for echoes \citep{serafinnadeau24}.

    In our new observations, several instances of echoes that do not reach zero delay can be readily identified; see Figure~\ref{fig:full_data}.
    These largely cluster around the period of low-curvature echoes in the latter half of 2022 (see Sect.~\ref{sec:low_curv}).
    As can be seen in the more detailed view in Figure~\ref{fig:nonzero},
    these echoes are extremely long-lived, evolving very slowly, with the longest lasting up to 130 days, and reach minimum delays of $\sim\!0.5{\rm\,ms}$.

    \cite{serafinnadeau24} used the absence of echoes that do not reach zero delay to set a lower limit to the lengths of the lensing structures, of $\gtrsim\!4{\rm\,au}$.
    At first glance, having now found non-crossing echoes, it seemed we should be able to use this for an actual estimate.
    This is not straightforward, however, as the geometry assumed in \citet[their Figure~6]{serafinnadeau24} is that the lensing structures are thin and long, with well-defined ends.
    For that case, it would be expected that while some echoes might not start or end at zero delay, they would still evolve at similar rates as those that do.
    Indeed, ideally, we would have some non-crossing echoes in multiple groups of echoes with similar curvature (and thus likely similar orientation).
    The non-crossing echoes we observe, however, evolve much more slowly than any other echo, with a curvature of only $\sim\!0.12{\rm\,\unit{\micro\second}/day^2}$.
    This very low curvature, combined with the absence of well-defined starts or ends (or at least changes in curvature), means we cannot take a ratio relative to similar echoes that do cross the line of sight, and hence cannot yet make an estimate of the lengths in this way.

    The evolution of these echoes, as with others, appears largely parabolic in delay, with arcs both approaching and receding from the pulsar.
    This similarity is however strange if these are also produced by linear structures.
    As the pulsar approaches a linear structure we expect a parabolic arc to one side.
    However, for the minimum delay to be nonzero, the structure can never directly intersect the line of sight, and so at closest approach, the echo is produced near the end of the filament, and the separation can then no longer evolve linearly with time.
    The delay that would be expected as the pulsar recedes from the structure would then no longer evolve quadratically, but differently, likely much more rapidly, depending on the specific geometry at the end of the filament.

    The overall properties of the non-crossing echoes, however, can be understood with a small modification of the above picture, in which the lensing structures are not straight but mildly curved on the sky (which may in fact be expected if, as suggested by \cite{serafinnadeau24}, they arise in the thin ionized skins of roughly cylindrical filaments).
    If so, the non-crossing echoes we see are due to the line of sight to the pulsar passing very close by such mildly curved structures, moving nearly parallel to it.
    In this case, their duration $\Delta t$ directly gives a lower limit to the structure's length, $L\gtrsim v_{ps}\Delta t\simeq9{\rm\,au}$ (for $\Delta t=130{\rm\,d}$ and $v_{\rm ps}=120{\rm\,km/s}$).

    Note that during this time, the relative magnification of the non-crossing echo still implies that the structure is thin ($W\simeq0.1{\rm\,au}$, \citealt{serafinnadeau24}), i.e., it is highly elongated.
    Also, the perpendicular offset between the structure and the pulsar remains small: the range in delay of 0.5 to $1.0{\rm\,ms}$ implies $x=0.45$ to $0.6{\rm\,au}$ (Eq.~\ref{eqn:bending}, assuming $d_{\rm ps}=0.5{\rm\,pc}$).
    An estimate, or perhaps more likely an upper limit, to the length of the structures may then be the radius of curvature, $R_c\simeq \frac12 (L/2)^2/\Delta x \simeq 70 {\rm\,au}$.

    Two final notes.
    First, the delay for the non-crossing echoes shows smooth variations.
    This could be due to multiple echoes contributing, but might also indicate deviations from  a quadratic dependence.
    If the latter, and if the pulsar is indeed passing close by the lensing structure, this would indicate that the structure has small-scale variations in locations or orientations.
    Second, for a lens structure that is not a straight line projected on the sky, the perpendicular distance between the pulsar and the lens will no longer change linearly (as assumed in Eq.~\ref{eqn:delay_evol}) and hence the geometric delay will no longer evolve exactly quadratically.
    Given the large radius of curvature, however, any deviations will be small for all but the most slowly evolving echoes, where the pulsar moves nearly parallel to the structure.

    \subsection{Chromatic Behaviour}\label{sec:chromatic}

    \begin{figure*}
        \centering
        \includegraphics[width=\textwidth]{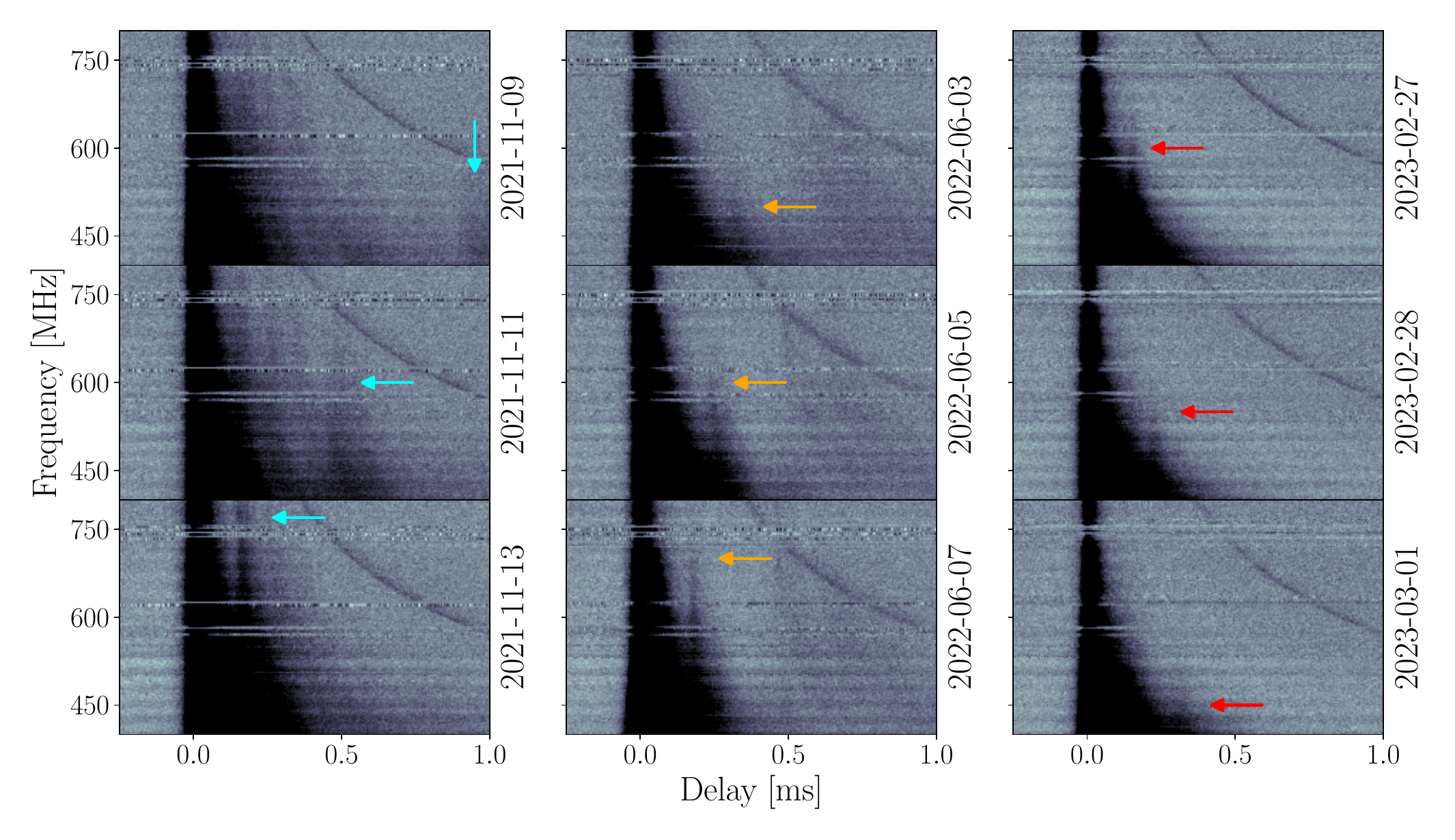}
        \caption{
            Three echoes with strong frequency evolution, visible out to much larger delays at lower frequencies.
            Unlike most echoes, for these three, the echo appears or disappears not by fading away at large delay, but by entering or exiting the bottom of the observation bandwidth.
            For each echo, we show the pulse stacks for 3 days:
            \textit{Left:\/} 2021 November 9, 11, and 13, showing an echo\review{ (cyan arrow)} that appears at large delays at low frequencies only, and then rapidly moves in.
            \textit{Middle:\/} 2022 June 3, 5, and 7, with an echo\review{ (orange arrow)} that just barely sticks out over the scattering tail, moving to higher frequencies as it moves in.
            \textit{Right:\/} 2023 February 27, 28, and March 1, with an echo\review{ (red arrow)} that similarly peeks out over the scattering tail, at relatively high frequency initially, but only out to lower frequencies as it moves to larger delay.
        }\label{fig:peekaboo}
    \end{figure*}

    Ionized material naturally bends lower frequency light more easily than light at higher frequencies, as can be seen from the frequency dependence of the bending angle, $\hat\alpha\propto\nu^{-2}$ (see Eq.~\ref{eqn:bending}).
    One would therefore expect that echoes will be visible for a wider range of delays as the observation frequency decreases.
    Indeed, combining with the dependence of geometric delay on bending angle (Eq.~\ref{eqn:delay}), the maximum delay out to which an echo can be seen should scale as~$\nu^{-4}$.

    In most echoes, no clear chromatic behaviour is seen; instead, they seem to simply appear or disappear at larger delays (see, e.g., the strong echoes in Figure~\ref{fig:overdense}).
    A few observed echoes, however, do follow the expectation; we show the clearest examples in Figure~\ref{fig:peekaboo}.

    A possible clue as to why only these echoes evolve chromatically may be that they are substantially narrower, i.e., span a smaller range of delay in a given instant, than most echoes in our dataset (especially the clearest cases, the second and third echo shown in Figure~\ref{fig:peekaboo}).
    In principle, a range in delay could be both intrinsic to the echo, if the echo were composed of multiple images arising in different parts of a lensing structure, or extrinsic, if a single image from one lensing structure was scattered further by other lensing structures.
    Given the variation in width of echoes, the cause must be at least partially intrinsic, with wider echoes likely having more images, spread over a larger area of the lensing structure.
    The frequency dependence will then enter only via the number of images: more will contribute at lower frequencies.

    If wider echoes indeed contain more images, that might explain why they do not show the expected chromatic behaviour: towards larger delays and thus bending angles, at each frequency some of the images will disappear, and the echo will just fade.
    In contrast, the chromatic echoes, for which the smaller range in delay implies that they arise over smaller areas of lensing structure, may have only one or a few images, so that the frequency dependence of a given image can more easily be noticed.
    This also suggests, perhaps counterintuitively, that weaker lenses produce the most sharply defined echoes, consistent with the fact that the sharp echoes discussed here are relatively weak.

    What could be the physical reason for the difference in number of images?
    \cite{serafinnadeau24} suggested that the lensing structures responsible for producing echoes are likely to be the thin, ionized skins of extended filamentary structures in the nebula, which are seen at grazing incidence.
    Since the filaments are in the process of being evaporated, their ionized skins are unlikely to be smooth, and it is not implausible that they lead to multiple images, surrounding the point where the unperturbed structure would produce an image.
    In the case of a stronger lens, the range over which perturbations are capable of producing images would be larger than for a weaker lens, and therefore the echo images from stronger lenses might be expected to be more broadened.
    In addition, the roughness of the lens may simply be variable, leading to a variable number of images, analogously to how, when one sees the Sun set over water, the number of reflected images will depend on the water's choppiness.

    \begin{figure}
        \centering
        \includegraphics[width=\columnwidth]{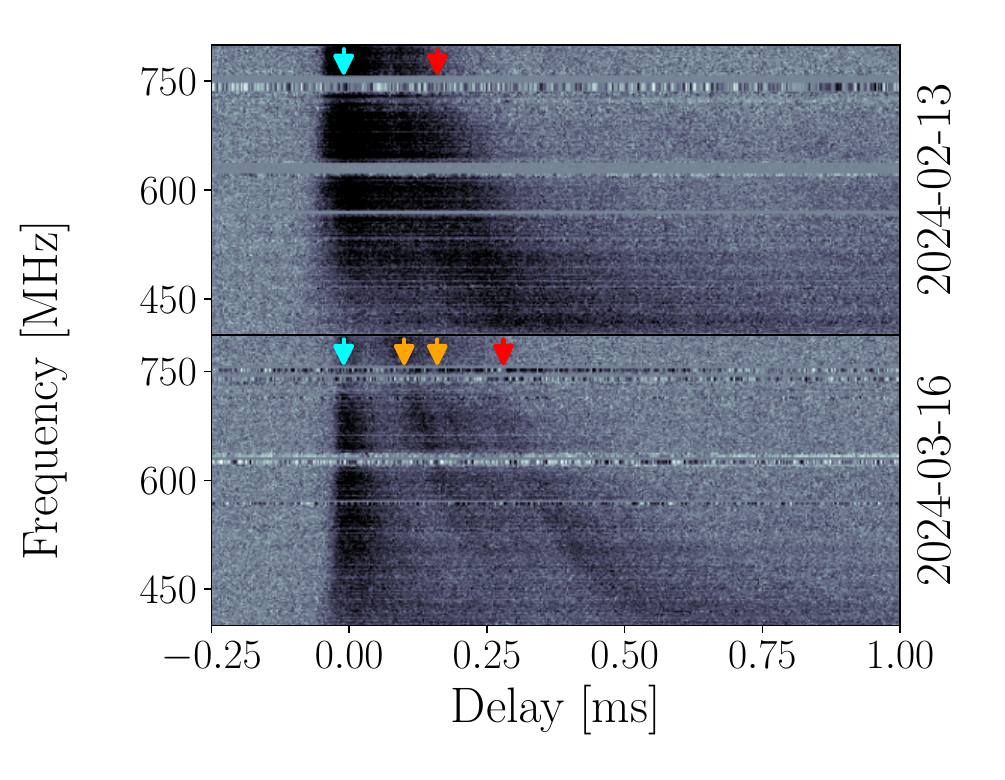}
        \caption{
            Daily stacks from 2024 February 13 \textit{(top)} and March 16 \textit{(bottom)}, both of which show complex frequency and delay dependent behaviour, which we suggest is due to demagnification at low frequencies (see Section~\ref{sec:chromatic}). \review{Identifiable components are highlighted with arrows: cyan for the primary components, and orange and red for echo components.}
        }\label{fig:frb}
    \end{figure}

    We now turn to a few additional cases of echoes that exhibit unusual frequency structure.
    We have noticed two kinds, and show an example of each in Figure~\ref{fig:frb}.
    For the first, seen on 2024 February 13, the echo\review{ (red arrow)} is visible across the observation bandwidth, while the primary component\review{ (cyan arrow)} is weak below $500{\rm\,MHz}$.
    This frequency structure can be understood if light from the primary component, which travels closest to the line of sight, crosses through an overdense lensing structure (likely the same that produces the echoes).
    Since bending angles are larger at lower frequencies, the low-frequency part may be bent out of the line of sight and thus demagnified, while its higher frequency emission can pass through comparatively unobstructed.

    In the second case, observed in a pair of echoes which are most visible on 2024 March 16 and 17\review{ (pair of orange arrows)}, the echoes are clearly visible at higher frequencies but very weak and difficult to see below $500{\rm\,MHz}$.
    Intriguingly, alongside both echoes receding away from the line of sight is a third\review{ (red arrow)}, at slightly larger delay, which is itself visible over the full observation bandwidth.
    Such a morphology is much more difficult to explain than in the prior case.
    We note, however, that this set of echoes was observed at the tail end of the active period of winter 2024, and it is clear from discussion of echoes during the 2022 March high-activity period (Sect.~\ref{sec:overdensity}) that echoes can be demagnified by subsequent images.
    Therefore, it may be that this unusual morphology is a consequence of the two intermediate echoes being demagnified at low frequencies by the more delayed echo, while the line of sight image\review{ (cyan arrow)} is sufficiently far moved from that overdensity to be comparatively unaffected.
    If so, this implies that the structures responsible for the more delayed echo must be located in front or behind of those responsible for the intermediate ones---they cannot be co-located.

\section{Ramifications}\label{sec:ramifications}

    In this paper, we presented a daily archive of Crab giant pulses which we use to study echoes over a period of two and a half years, between late 2021 and early 2024.
    To do so, we expand the alignment stacking method presented by \cite{serafinnadeau24} to \ac{CHIME}'s wide bandwidth and show that this produces profile averages which significantly improve on folded data, which smear out giant pulses and cause echoes to blend into the profile's scattering tail.
    Indeed, as was the case with the archival Jodrell Bank data searched by \cite{serafinnadeau24}, we note that echoes are readily visible throughout our dataset using this method.
    The consistency with which these events have been observed over such long time periods demonstrates that the Crab Nebula is an excellent laboratory for studying plasma lensing.
    When echoes are not present, the scattering is low and varies very little.
    This implies that material associated with echoes dominates the nebular contribution of scattering, and scattering that remains in the absence of this material is likely dominated by the interstellar medium, as suggested by \citet{rudnitskii16} and \citet{rebecca23b}.

    Throughout these observations, we see discrete shifts in arc curvatures between groups of echoes.
    We therefore conclude that echoes in the Crab Nebula must be the product of material associated with larger structures which constrain the distances, velocities and orientations of lenses for a cluster of echoes.
    Among these clusters, we report on a group of extremely low curvature ($\sim 1 {\rm\,\unit{\micro\second}/day^2 }$) echoes through much of 2023, which evolve even more slowly than the 1997 event \citep[$2 {\rm\,\unit{\micro\second}/day^2 }$][]{lyne01}.
    Furthermore, we use the highest curvature echoes ($\eta = 20{\rm\, \unit{\micro\second} / day^2}$) to constrain the velocity of structures close to the line of sight, with $v_{\rm eff} = 157 {\rm\, km/s}$.

    We also identify two clusters of echoes, around 2022 March and 2024 March, which are characterized by large dispersion measure changes as the profile becomes dominated by echo images, and by demagnification as the lensed images come near the line of sight, consistent with overdense lenses.
    The simultaneous existence of multiple images at different \acp{DM} shows that the profile dispersion measure cannot truly trace the column density across the lens and therefore modeling echoes by simply using the observed \ac{DM} curve as a proxy for the lens column density will not necessarily be reflective of the actual structures involved.

    Even outside of these two clusters, we see that echoes appear largely consistent with lensing by overdensities, though the dispersive delays and the resulting demagnification are often far less pronounced if visible.
    We do however see two echoes (in 2022 May and 2023 January) with apparent deficits in dispersion, resulting in delays which decrease at lower frequencies.
    The sources of underdensities in the nebula are not obvious, unlike the many overdense regions which have been observed \citep{osterbrock57}.
    We note however that it is nonetheless possible that these echoes may instead be the result of narrow gaps between overlapping overdense structures.

    Additionally, we show the existence of echoes which never directly cross the line of sight.
    These were expected to occur, but had not been observed since the discovery of echoes, over 20 years ago.
    In principle, echoes like these may allow one to estimate typical lengths of the lensing structures based on their rate of occurrence relative to the more commonly seen line of sight crossing echoes.
    However, such estimate are made suspect by the fact that the curvatures are much lower than those of other echoes.
    Instead, as the low curvature indicates the pulsar travels nearly parallel to the structure, we use the observed timescale to set a lower limit of $9\,\unit{au}$.
    It may be possible to obtain better estimates for the filament lengths from continued monitoring of the Crab for echoes and identifying clusters which have both line of sight crossing and non-crossing events, and by keeping track over time of the number of clusters with given curvatures  and the relative number of clusters with and without non-crossing echoes.

    Although some echoes are morphologically simple, others across the observation period discussed in this paper display complex frequency and delay behaviour.
    The frequency structure observed in these examples, as well as many other examples primarily focused around the high activity periods described in Section \ref{sec:active}, are strikingly similar to a wide array of fast radio bursts (FRBs) with more complex morphologies, during which multiple bursts are observed \citep{faber24, sand25, tian25}, and in particular the two cases shown in Fig.~\ref{fig:frb} are similar to examples of ``upward drifting'' and ``downward drifting'' FRBs.
    Indeed, it has been shown that some FRBs inhabit ionized environments \citep{michilli18a}, and it has previously been suggested that these complex morphologies may be caused by plasma lensing \citep{platts21, faber24} similar to that observed here \citep{cordes17, hessels19}.

\vspace{5mm}
\textit{Acknowledgements}:
We thank the Toronto scintillometry group for useful discussions.
\review{We also thank the anonymous referee for their review, which helped clarify a number of passages throughout the paper.}
We acknowledge that \ac{CHIME} is located on the traditional, ancestral, and unceded territory of the Syilx/Okanagan people. We are grateful to the staff of the Dominion Radio Astrophysical Observatory, which is operated by the National Research Council of Canada. CHIME operations are funded by a grant from the NSERC Alliance Program and by support from McGill University, University of British Columbia, and University of Toronto. CHIME was funded by a grant from the Canada Foundation for Innovation (CFI) 2012 Leading Edge Fund (Project 31170) and by contributions from the provinces of British Columbia, Québec and Ontario. The CHIME/FRB Project, which enabled development in common with the CHIME/Pulsar instrument, was funded by a grant from the CFI 2015 Innovation Fund (Project 33213) and by contributions from the provinces of British Columbia and Québec, and by the Dunlap Institute for Astronomy and Astrophysics at the University of Toronto. Additional support was provided by the Canadian Institute for Advanced Research (CIFAR), the Trottier Space Institute at McGill University, and the University of British Columbia. The CHIME/Pulsar instrument hardware is funded by the Natural Sciences and Engineering Research Council (NSERC) Research Tools and Instruments (RTI-1) grant EQPEQ 458893-2014.
M.Hv.K. is supported by the Natural Sciences and Engineering Research Council of Canada (NSERC) via discovery and accelerator grants, and by a Killam Fellowship.
K.W.M. holds the Adam J. Burgasser Chair in Astrophysics and recieved support from an NSF grant (2018490).
U.P.  is supported by the Natural Sciences and Engineering Research Council of Canada (NSERC), [funding reference number RGPIN-2019-06770, ALLRP 586559-23],  Canadian Institute for Advanced Research (CIFAR), AMD AI Quantum Astro.
Computations were performed on the Sunnyvale computer at the Canadian Institute for Theoretical Astrophysics (CITA).

\vspace{5mm}
\facilities{CHIME \citep{chime22}.}

\vspace{5mm}
\software{astropy \citep{astropy13, astropy18, astropy22},
  baseband \citep{baseband25},
  baseband-tasks \citep{mhvk25},
  matplotlib \citep{matplotlib07},
  numpy \citep{numpy20},
  scipy \citep{scipy20}.}

\bibliography{main}{}
\bibliographystyle{aasjournal}

\end{document}